\documentclass[superscriptaddress,aps]{revtex4} 
\usepackage{epsfig}
\usepackage{verbatim}
\usepackage{array}

\newcommand{\be}{\begin{equation}}
\newcommand{\ee}{\end{equation}}
\newcommand{\ba}{\begin{eqnarray}}
\newcommand{\ea}{\end{eqnarray}}

\newcommand{\bra}{\langle }
\newcommand{\ket}{\rangle}
\usepackage{graphicx}
\usepackage{subfigure}
\usepackage{color} 
\usepackage{amsfonts}
\usepackage{amsmath}
\usepackage{appendix}
\usepackage{lipsum}

\begin{document}
\title{Semiclassical fast-forward shortcuts to adiabaticity}
\author{Ayoti Patra}
\affiliation{Department of Physics, University of Maryland, College Park, Maryland 20742, USA}
\author{Christopher Jarzynski}
\affiliation{Department of Chemistry and Biochemistry, University of Maryland, College Park, Maryland 20742, USA}
\affiliation{Institute for Physical Science and Technology, University of Maryland, College Park, Maryland 20742, USA}
\affiliation{Department of Physics, University of Maryland, College Park, Maryland 20742, USA}

\begin{abstract}
In {\it fast forward} quantum shortcuts to adiabaticity, a designed potential $U_{FF}(q,t)$ steers a wavefunction to evolve from the $n$'th eigenstate of an initial Hamiltonian $\hat H(0)$ to the $n$'th eigenstate of a final Hamiltonian $\hat H(\tau)$, in finite time $\tau$.
Previously proposed strategies for constructing $U_{FF}$ are (in the absence of special symmetries) limited to the ground state, $n=0$.
We develop a method that overcomes this limitation, thereby substantially expanding the applicability of this shortcut to adiabaticity, and we illustrate its effectiveness with numerical simulations.
Semiclassical analysis provides insight and establishes a close correspondence with the analogous classical fast forward method.
\end{abstract}
\maketitle

\section{Introduction}

Shortcuts to adiabaticity (STA) -- strategies for attaining adiabatic outcomes in finite times -- offer a set of tools for accelerating the dynamics of quantum systems,
with important applications to atomic and chemical physics and quantum technologies \cite{Torrontegui13,Masuda16,Guery-Odelin19}.
Considerable theoretical \cite{Rice05,Berry09,Masuda10,Masuda11,Ibanez11,Chen11,Torrontegui12,Ibanez12,dCampo13,Torosov13, Vacanti14, Torrontegui14,Martinez14,Deffner14,Takahashi15,Song16,Deffner16,Muga16,Martinez-Garaot16,Mukherjee16,Song17,Sels17} and experimental \cite{Schaff10,Schaff11,Bason11, Walther12,Zhang13,Kihwan16,Wang18} advances have been made in this emerging field, and have been extended to classical Hamiltonian \cite{Jarzynski13,Deng13,Patra16, Okuyama16, Jarzynski17,Torrontegui17} and stochastic dynamics \cite{Tu14,Martinez16,Li17,Patra17,Chupeau18}.

For continuous systems with one degree of freedom, we formulate the problem of STA as follows (see Sec.~\ref{sec:conclusions} for a slightly different formulation).
Consider a Hamiltonian
\be
\label{eq:H0}
\hat H_0(t) = -\frac{\hbar^2}{2m} \frac{\partial^2}{\partial q^2} + U_0(q,t)
\ee
and let $\vert k(t)\rangle$ or $\phi_k(q,t)=\langle q\vert  k(t)\rangle$ denote its $k$'th eigenstate, with eigenvalue $E_k(t)$. 
Here $\phi_k(q,t)$ is a real function and $\hat H_0(t)$ is time-dependent only in the interval $0\le t\le\tau$.
Initializing the system in an energy eigenstate $\psi_{t\le 0} = \vert n(0)\rangle$,
we want it to evolve to the final eigenstate $\psi_{t\ge\tau} = \vert n(\tau)\rangle$.
For slow driving ($\tau\rightarrow\infty$) the system naturally follows the adiabatic path $\psi_t=\vert n(t)\rangle$ at all times~\cite{Griffiths04}.
But for rapid driving, additional measures -- ``shortcuts'' -- must be taken to prevent non-adiabatic transitions and guide the system to the desired final state.
Here and below we suppress overall time-dependent phases when writing the state of our system.

In one STA approach, known as {\it fast-forward} (FF) driving~\cite{Masuda10,Masuda11,Torrontegui12}, an auxiliary potential $U_{FF}(q,t)$ is added to $\hat H_0(t)$, where $U_{FF}$ is designed to guide the system to the final state $\psi_\tau=\vert n(\tau)\rangle$.
This approach offers a promising tool for achieving the controlled acceleration of continuous quantum systems, towards desired ends such as 
the manipulation of Bose-Einstein condensates~\cite{Torrontegui12a,Masuda12,Masuda14,Masuda14a,Masuda18} or ions in trapping potentials~\cite{Masuda15,Kiely15}, or population transfer between vibrational molecular states~\cite{Masuda15a,Masuda16}.

Unfortunately the procedure for constructing $U_{FF}(q,t)$ generically leads to divergent behavior -- ``infinities'' in the potential -- at locations where the eigenstate $\phi_n(q,t)$ has nodes~\cite{Martinez-Garaot16,Patra17,Guery-Odelin19}.
As a result this method has been restricted to ground-state wavefunctions, which lack nodes, or else to situations in which special scale-invariant symmetries eliminate the divergences \cite{Deffner14}.
Truncating divergences with finite cutoffs has proven useful when driving a system from its ground state to its first excited state (i.e.\ $\vert 0(0)\rangle \rightarrow \vert 1(\tau)\rangle$)~\cite{Martinez-Garaot16},
but this strategy has not been explored for the STA problem formulated above ($\vert n(0)\rangle \rightarrow \vert n(\tau)\rangle$), and it is likely to lead to difficult-to-engineer potentials for eigenstates with many nodes.

In this paper we extend the fast-forward approach to excited states, by developing a semiclassical strategy that is free of divergences.
In Sec.~\ref{sec:main} we derive our main result: a semiclassically motivated recipe for constructing a divergence-free $U_{FF}(q,t)$ for excited energy eigenstates.
In Sec.~\ref{sec:numerical} we use numerical simulations to illustrate the effectiveness of this approach.
We find that our fast-forward potential steers the wavefunction to the desired final eigenstate with high accuracy, with only a small amount of probability spilling out into sideband states.
In Sec.~\ref{sec:classical} we compare these results with corresponding classical simulations, and we develop a semiclassical theory for these sidebands.
We end with a brief discussion in Sec.~\ref{sec:conclusions}.
Several technical steps of our analysis are relegated to the Appendix, for clarity of presentation.

\section{Derivation of main result}
\label{sec:main}

As in Refs.~\cite{Jarzynski17,Patra17}, our starting point is a velocity field $v(q,t)$ (as yet undetermined) that vanishes for $t \notin [0,\tau]$.
We then use $v(q,t)$ to define an acceleration field $a(q,t)$ from which the fast-forward potential is constructed, as follows:
\begin{equation}
\label{eq:aUFF}
\partial_q U_{FF}(q,t) = -ma(q,t) \equiv -m ( v \partial_q v + \partial_t v )
\end{equation}
Given $v(q,t)$, Eq.~\ref{eq:aUFF} defines $U_{FF}(q,t)$ up to an integration constant $u_{FF}(t)$, which is chosen so that $U_{FF}=0$ for $t\notin[0,\tau]$.
Thus $v$, $a$ and $U_{FF}$ all vanish outside the interval $[0,\tau]$.

The task now is to design $v(q,t)$ so that $U_{FF}(q,t)$ (given by Eq.~\ref{eq:aUFF}) generates the desired shortcut.
To this end we define a function $S(q,t)$ via the relation
\ba
\label{eq:S}
\partial_q S(q,t) = mv(q,t)
\ea
where the integration constant $s(t)$ is adjusted (see Appendix for details) so that $S$ obeys the Hamilton-Jacobi equation
\be
\label{eq:HJ}
\frac{\partial S}{\partial t} + \frac{1}{2m} \left( \frac{\partial S}{\partial q} \right)^2 + U_{FF} = 0
\ee
By Eq.~\ref{eq:S}, $S(q,t) = S_-$ for $t\le 0$ and $S(q,t) = S_+$ for $t\ge \tau$, for some constants $S_\pm$.

With these definitions in place we propose the {\it ansatz}
\be
\label{eq:ansatz}
\bar\psi(q,t) = \phi_n(q,t) e^{iS(q,t)/\hbar} \exp \left[ -\frac{i}{\hbar} \int_0^t E_n(t^\prime) dt^\prime \right] 
\ee
(for a given $n\ge 0$) as a solution to the time-dependent Schr\" odinger equation (TDSE)
\be
\label{eq:tdse}
i\hbar\frac{\partial\bar\psi}{\partial t} = (\hat H_0  + \hat U_{FF})\bar\psi
\ee
Note that $\bar\psi_{t\le 0} = \vert n(0)\rangle$ and $\bar\psi_{t\ge\tau} = \vert n(\tau)\rangle$.
Substituting Eq.~\ref{eq:ansatz} into Eq.~\ref{eq:tdse} yields (see Appendix)
\be
\label{eq:SED}
i\hbar\frac{\partial\phi_n}{\partial t} = \frac{1}{2} \left ( \hat p \hat v + \hat v \hat p \right) \phi_n \equiv \hat D(t) \phi_n
\ee
where $\hat p = -i\hbar \partial_q$ and $\hat v(t) = v(\hat q,t)$.
Thus if we can find a function $v(q,t)$ that satisfies Eq.~\ref{eq:SED}, then by construction our {\it ansatz} $\bar\psi$ (Eq.~\ref{eq:ansatz}) will satisfy the TDSE, and will have the desired property of beginning and ending in the $n$'th energy eigenstate.
In other words
our {\it ansatz} converts the problem of constructing a potential $U_{FF}$ that satisfies Eq.~\ref{eq:tdse}, to that of constructing a velocity field $v$ that satisfies Eq.~\ref{eq:SED}, for a given choice of $n$.
We now explore strategies for accomplishing this task.
In what follows, we drop the subscript $n$, and simply write $\phi(q,t)$ and $E(t)$ for the $n$'th eigenstate and eigenenergy of $\hat H_0(t)$.

Eq.~\ref{eq:SED} implies (see Appendix)
\be
\label{eq:continuity}
\partial_t \phi^2 + \partial_q(v\phi^2)=0
\ee
which is the continuity equation for a probability density $\phi^2(q,t)$ evolving under the deterministic flow $dq/dt = v(q,t)$.
Thus it seems we should construct $v(q,t)$ to satisfy Eq.~\ref{eq:continuity}.
Unfortunately the resulting velocity field generically diverges at the nodes of the eigenstate, where $\phi^2=0$~\cite{Patra17}.
In fact, constructing $U_{FF}$ via Eq.~\ref{eq:continuity} is equivalent to previously proposed fast-forward approaches~\cite{Masuda10,Masuda11,Torrontegui12}, and leads to the  divergent behavior mentioned earlier.
Let us therefore try a different strategy, motivated by semiclassical arguments.

In the WKB approximation, an excited eigenstate of $\hat H_0(t)$ is given by~\cite{Griffiths04}
\be
\label{eq:wkb}
\phi(q,t) = \sqrt{\frac{\rho}{2}} \left( e^{i\Sigma/\hbar - i\pi/4} +  e^{-i\Sigma/\hbar + i\pi/4} \right)
\equiv \phi_+ + \phi_-
\ee
where
\be
\label{eq:Sigma}
\Sigma(q,t) =  \int_{q_L}^q \bar p(q^\prime,t) \, dq^\prime
\quad,\quad
\bar p = \sqrt{ 2m(E - U_0)}
\quad,\quad
\rho(q,t)\propto\frac{1}{\bar p}
\ee
Here we treat $t$ as a fixed parameter, and we consider a classical trajectory of energy $E = E(t)$ evolving under the Hamiltonian $H_0 = p^2/2m+U_0(q,t)$.
Then $q_L(t)$ is the left turning point of the trajectory; $\pm\bar p(q,t)$ are the momenta on the upper ($+$) and lower ($-$) branches of the {\it energy shell} (i.e. level set) $H_0(q,p,t)=E(t)$; and $\rho(q,t)$ is the position probability density sampled by the trajectory over one period of oscillation.
In Eq.~\ref{eq:Sigma} and below, we replace the eigenenergy $E$ by its WKB approximation, determined by $\oint_E \bar p \, dq = 2\pi\hbar [n+(1/2)]$.

\begin{figure}
\centering
\includegraphics[width=0.35\textwidth]{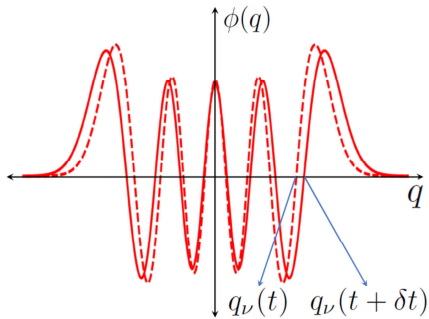}
\caption{An energy eigenstate $\phi$ at times $t$ and $t+\delta t$ (dashed and solid curves).
By Eq.~\ref{cond_real}, the node $q_\nu(t)$ is displaced by $v(q_\nu,t) \, dt$ when $t \rightarrow t+dt$.
}
\label{fig:nodes}
\end{figure}

The eigenstate probability distribution $\phi^2(q,t)$ oscillates with $q$ due to interference between $\phi_+$ and $\phi_-$ (Eq.~\ref{eq:wkb}), but after local averaging over these oscillations, $\phi^2(q,t)$ is semiclassically equivalent to $\rho(q,t)$~\cite{Griffiths04}.
Since the troublesome nodes in $\phi$ arise from destructive interference between $\phi_+$ and $\phi_-$, let us see what happens if we attempt to construct $v(q,t)$ to satisfy Eq.~\ref{eq:SED} separately for each term: $i\hbar \partial_t \phi_\pm = \hat D \phi_\pm$.
Upon substitution and separation of real and imaginary parts (see Appendix) we obtain
\begin{subequations}
\label{eq:perfect_cond}
\ba
\partial_t \Sigma + v \partial_q \Sigma &=&0 \label{cond_real} \\
\partial_t \rho + \partial_q(v \rho) &=& 0 \label{cond_im}
\ea
\end{subequations}
To interpret these conditions, we observe that $\phi_\pm(q,t)$ reflects both a classical probability distribution $\rho(q,t)$ and a quantum phase 
$e^{\pm i(\Sigma/\hbar - \pi/4)}$.
Eq.~\ref{cond_im} is a continuity equation for $\rho(q,t)$ under the flow $dq/dt = v(q,t)$, whereas
Eq.~\ref{cond_real} implies that the phase remains constant, $d\Sigma/dt=0$, under this flow.
By Eq.~\ref{eq:wkb} the nodes $\{ q_\nu(t) \}$ of $\phi(q,t)$ satisfy
\be
\Sigma(q_\nu(t),t) = \left( \nu - \frac{1}{4} \right) \pi\hbar \quad,\quad 1\le \nu\le n
\label{eq:preserve}
\ee
thus the value of $\Sigma$ remains constant along a node $q_\nu(t)$, as the parameter $t$ is varied.
Hence Eq.~\ref{cond_real} implies that the field $v(q,t)$ describes the nodal ``velocities'' (with respect to $t$), as illustrated in Fig.~\ref{fig:nodes}.

Eqs.~\ref{cond_real} and \ref{cond_im} are not generally satisfied by a single $v(q,t)$.
We have thus replaced one condition, Eq.~\ref{eq:continuity}, that suffers from divergences, with a pair of conditions, Eq.~\ref{eq:perfect_cond}, that cannot be simultaneously satisfied.
To make further progress, we choose to satisfy Eq.~\ref{cond_real} rather than Eq.~\ref{cond_im}.
Thus we take
\be
\label{eq:vqt}
v(q,t) = -\frac{\partial_t \Sigma}{\partial_q \Sigma} 
\ee
and construct $U_{FF}$ using Eq.~\ref{eq:aUFF}, in the hope that this fast-forward potential will accurately steer the wavefunction to the desired final adiabatic state.
{\it A posteriori} justification of this choice will be provided later.
First we describe a test of this approach, using numerical simulations of a model system evolving under the time-dependent Schr\" odinger equation.

\section{Numerical results}
\label{sec:numerical}

\begin{figure*}
   \begin{center}
      \subfigure{
         \label{fig:wocd_init}
         \includegraphics[trim = 2in 1in 2in 0in , scale=0.18,angle=0]{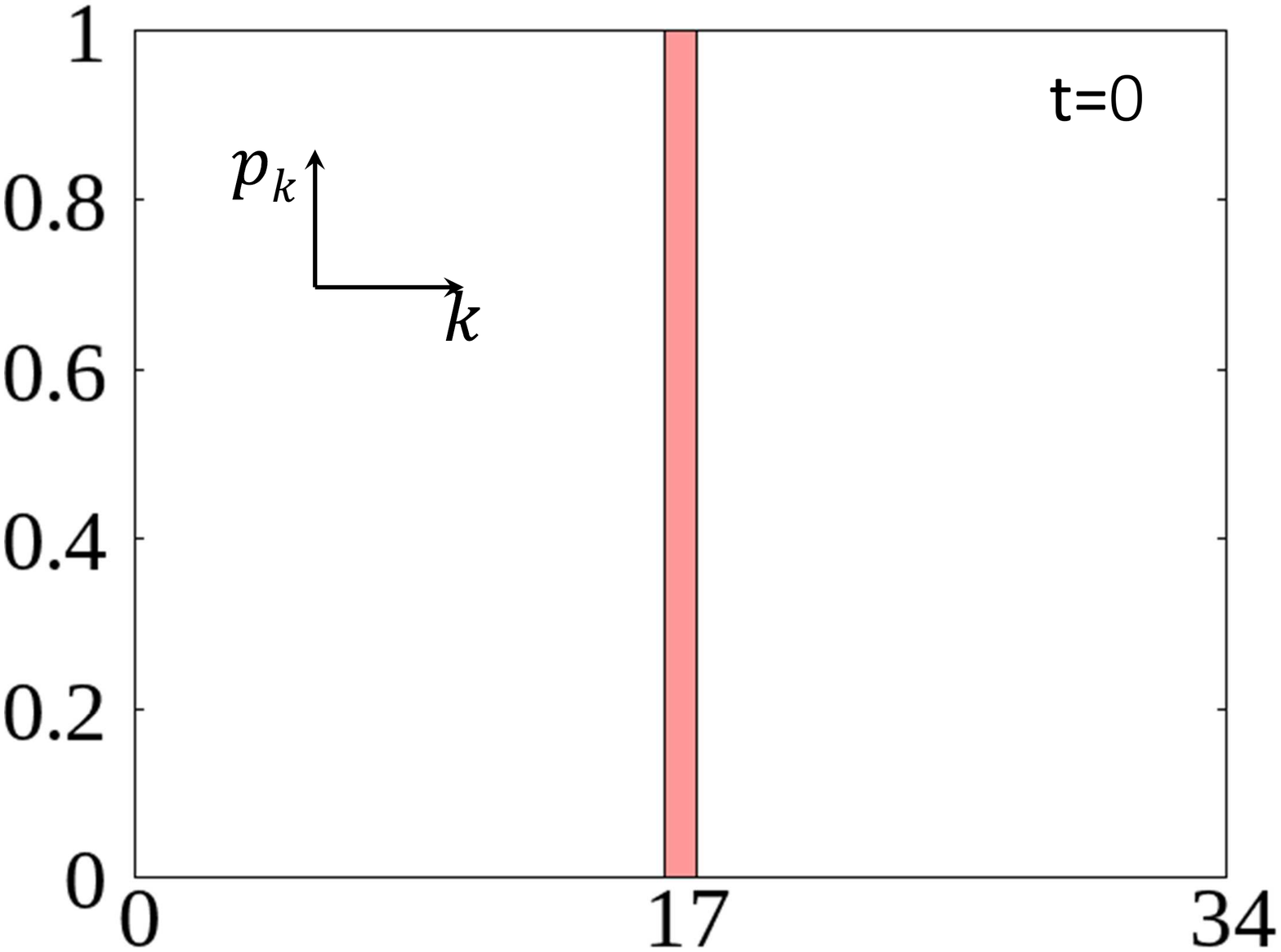}
      }
      \subfigure{
         \label{fig:wocd_interm}
         \includegraphics[trim = 0in 1in 2in 0in , scale=0.18,angle=0]{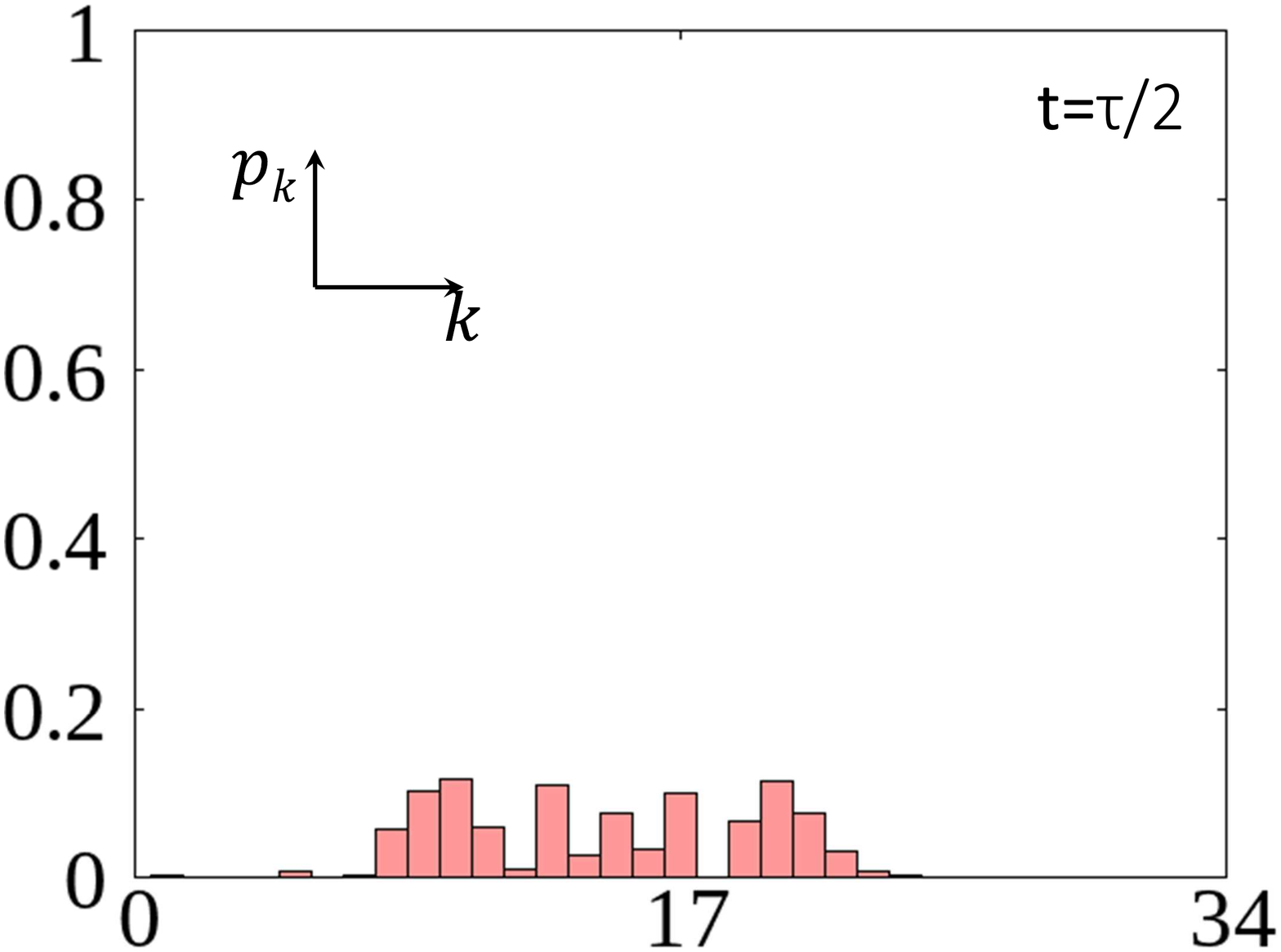}
      }
      \subfigure{
         \label{fig:wocd_final}
         \includegraphics[trim = 0in 1in 2in 0in , scale=0.18,angle=0]{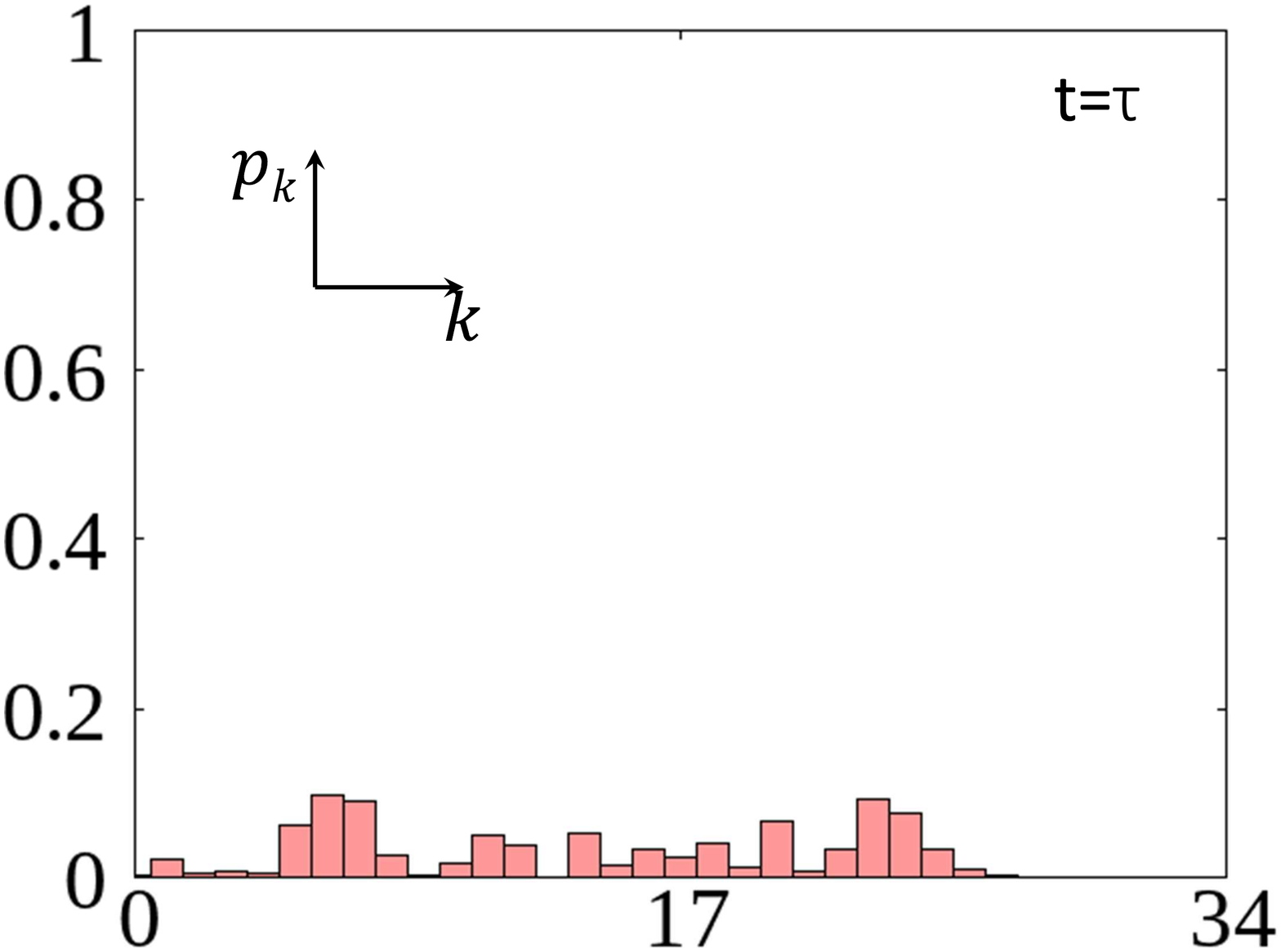}
      } \\
      \subfigure{
         \label{fig:wcd_init}
         \includegraphics[trim = 2in 1in 2in 0in , scale=0.18,angle=0]{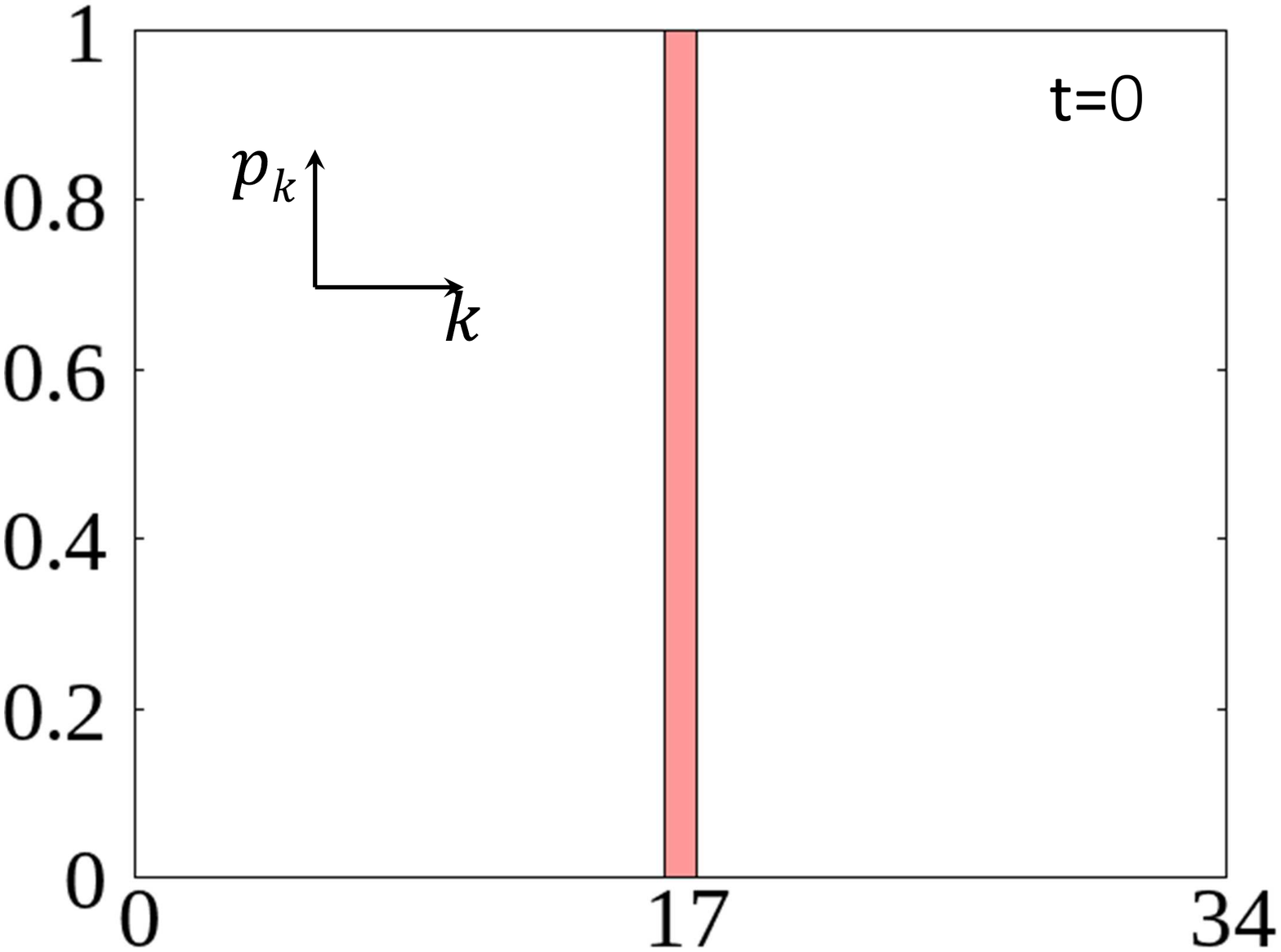}
      }
      \subfigure{
         \label{fig:wcd_interm}
         \includegraphics[trim = 0in 1in 2in 0in , scale=0.18,angle=0]{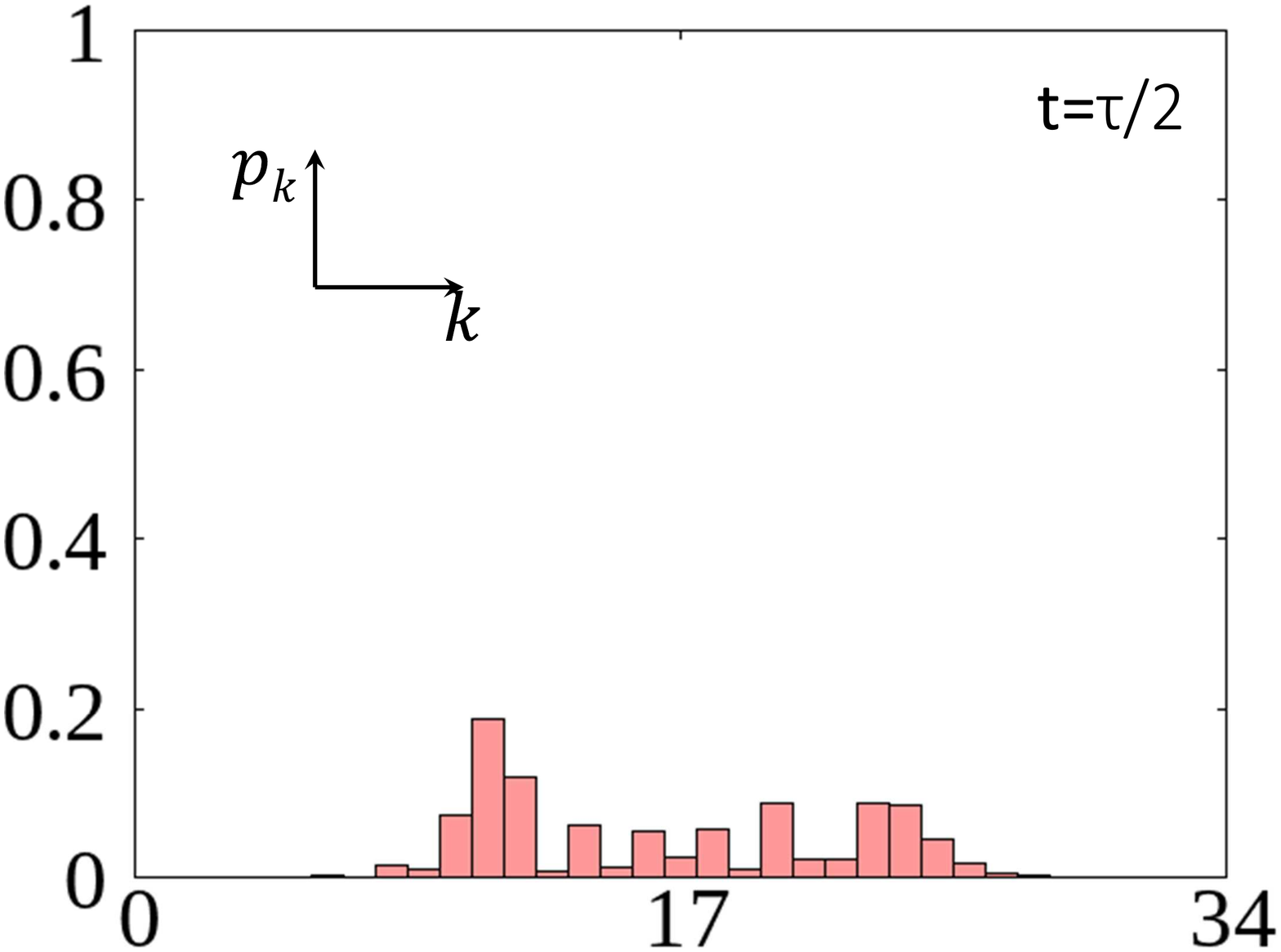}
      }
      \subfigure{
         \label{fig:wcd_final}
         \includegraphics[trim = 0in 1in 2in 0in , scale=0.18,angle=0]{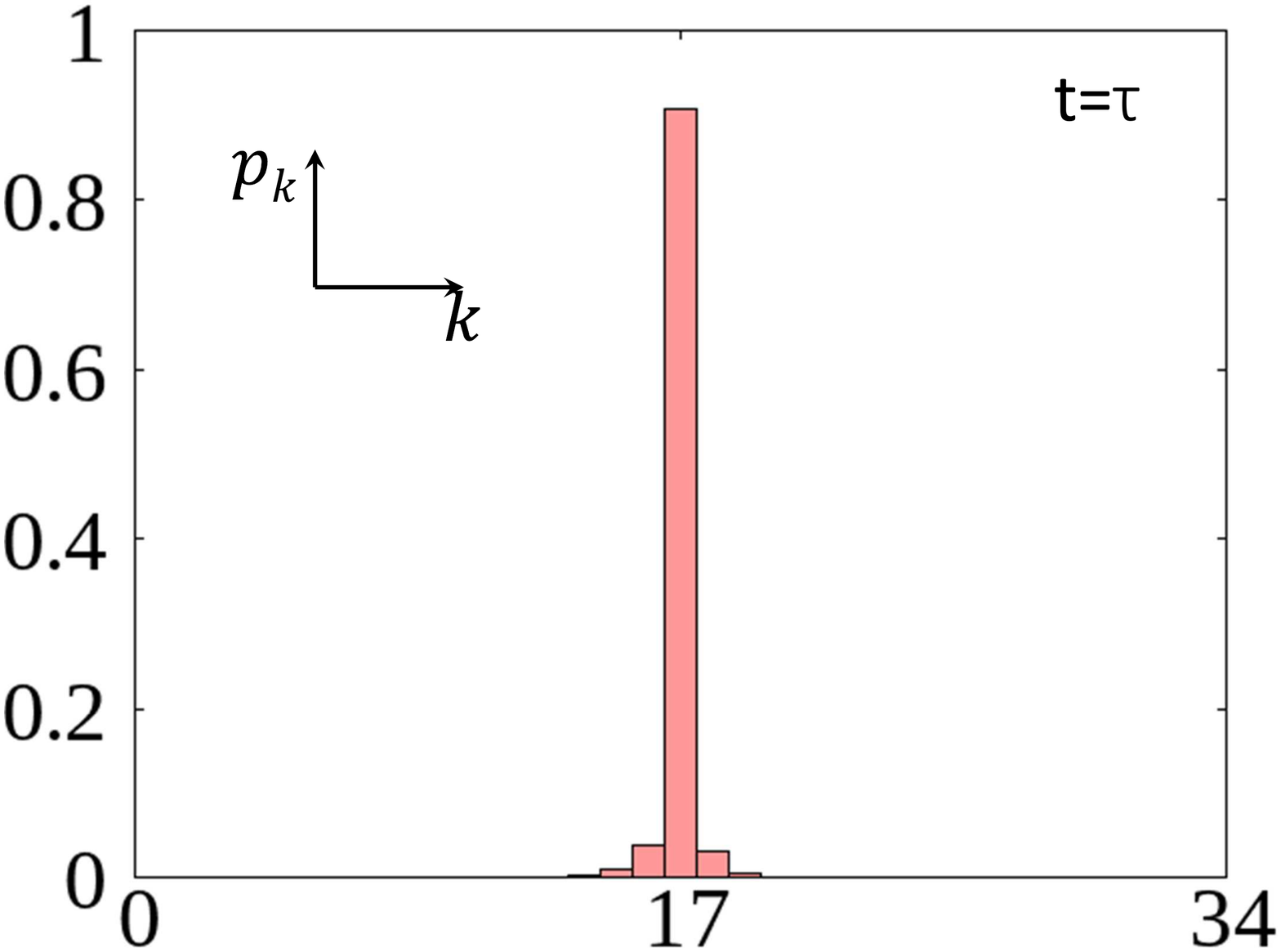}
      }
   \end{center}
   \caption{These plots show the overlap $p_k(t) = \vert \bra \phi_k(t) | \psi(t) \ket \vert^2$ between the eigenstates of $\hat H_0(t)$ and the wavefunction $\psi(t)$, as it evolves under $\hat H_0(t)$ (upper row) or $\hat H_0(t) + \hat U_{FF}(t)$ (lower row). 
   }
      \label{fig:prob_snapshot}
\end{figure*}

We simulated a quantum particle of unit mass evolving in a potential
\be
U_0(q,t) = q^4 - 16 q^2 + \lambda(t) q,
\label{eq:modelHamiltonian}
\ee
where $\lambda$ varies from $+16$ to $-16$ as $\lambda(t) = 4\cos(\pi t/\tau)[5- \cos(2\pi t/\tau)]$, with $\tau=1.0$. The system was initialized in the state $\psi_0=\vert n(0)\rangle$, and then evolved under the time-dependent Schr\"odinger equation -- first using the Hamiltonian $\hat{H}_0(t)$, then using $\hat{H}_0(t)+\hat U_{FF}(t)$, with the fast-forward potential determined by Eqs.~\ref{eq:aUFF} and \ref{eq:vqt} as described above.

Fig.~\ref{fig:prob_snapshot} presents numerical results for $n=17$, setting $\hbar=2$.
Each frame displays a histogram showing how the evolving wavefunction $\psi_t$ is distributed among the instantaneous eigenstates $|\phi_k(t)\ket$, as quantified by the overlap $p_k(t)=|\bra \phi_k|\psi_t \ket|^2$.
Figs.~\ref{fig:wocd_init} - \ref{fig:wocd_final} and \ref{fig:wcd_init} - \ref{fig:wcd_final} correspond to evolution under $\hat H_0$ and $\hat H_0+\hat U_{FF}$, respectively.
In both cases the system evolves from $\psi_0 = | 17 \ket$ (Figs.~\ref{fig:wocd_init}, \ref{fig:wcd_init}) to a superposition of energy eigenstates at intermediate times (Figs.~\ref{fig:wocd_interm}, \ref{fig:wcd_interm}).
While the system evolving under $\hat{H}_0(t)$ ends in a broad superposition of eigenstates, Fig.~\ref{fig:wocd_final}, the one evolving with the fast-forward shortcut reaches the final adiabatic state to a very good approximation: in Fig.~\ref{fig:wcd_final}, $p_{17}=0.91$ and $p_{16}+p_{17}+p_{18}=0.98$. 
Thus the potential $U_{FF}(q,t)$ constructed from Eqs.~\ref{eq:aUFF} and \ref{eq:vqt} achieves fast-forward driving with high accuracy.
We have performed simulations using different parameter values, and have found that the potential $\hat U_{FF}$ consistently guides the system to a final state whose probability is concentrated in the target eigenstate $\vert n(\tau)\rangle$, along with a few sideband states such as $k=16,18$ in Fig.~\ref{fig:wcd_final}.
(See Appendix for movies showing the evolving profile of $p_k(t)$ both without and with the fast forward potential.)

\begin{figure}
\centering
\includegraphics[trim = 0in 1.8in 0in 1in , width=0.5\textwidth]{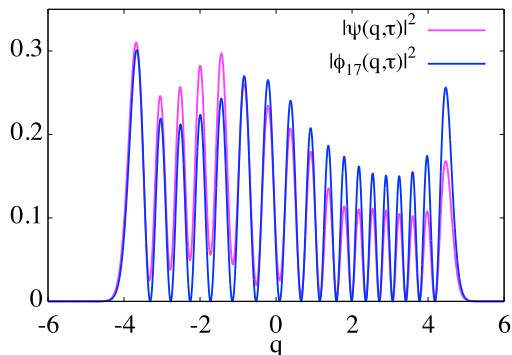}
\caption{For evolution under $\hat H_0+\hat U_{FF}$ from the initial state $\psi(q,0) = \phi_{17}(q,0)$, the final probability distribution $\vert\psi(q,\tau)\vert^2$ (magenta) and eigenstate distribution $\vert \phi_{17}(q,\tau)\vert^2$ (blue) are plotted.
}
\label{fig:compareProbDists}
\end{figure}

For evolution under $\hat H_0+\hat U_{FF}$, Fig.~\ref{fig:compareProbDists} shows the distributions $\vert\psi(q,\tau)\vert^2$ and $\vert \phi_{17}(q,\tau)\vert^2$.
While the minima and maxima align nicely, the values of $\vert\psi\vert^2$ and $\vert \phi_{17}\vert^2$ differ visibly.
This is understandable, as $v(q,t)$ was constructed to track the eigenstate's phase $\Sigma$ (Eq.~\ref{cond_real}) rather than its magnitude $\rho$ (Eq.~\ref{cond_im}).
The mismatch in Fig.~\ref{fig:compareProbDists} is reflected in the sidebands in Fig.~\ref{fig:wcd_final}.
We now develop intuition by comparing our results with classical simulations, and we derive a semiclassical prediction for the sidebands (Eq.~\ref{eq:fidelity}).

\section{Classical results and semiclassical analysis}
\label{sec:classical}

We first note that the potential $U_{FF}(q,t)$ determined by Eqs.~\ref{eq:aUFF} and \ref{eq:vqt} is identical to the one designed in Ref.~\cite{Jarzynski17} (see Eqs. 8-12 therein) for a purely classical fast-forward shortcut.
This agreement reflects the Correspondence Principle (CP) and provides a measure of justification for choosing to satisfy Eq.~\ref{cond_real} rather than Eq.~\ref{cond_im}.
Indeed, by invoking the CP we could have argued for constructing $U_{FF}(q,t)$ from Eqs.~\ref{eq:aUFF} and \ref{eq:vqt} directly from the classical results of Ref.~\cite{Jarzynski17}, rather than by proceeding via the WKB approximation as done above.
These considerations further motivate us to compare our quantum simulations with classical counterparts.

\begin{figure}
   \begin{center}
      \subfigure{
         \label{fig:class_init}
         \includegraphics[trim = 2in 1in 2in 0in , scale=0.18,angle=0]{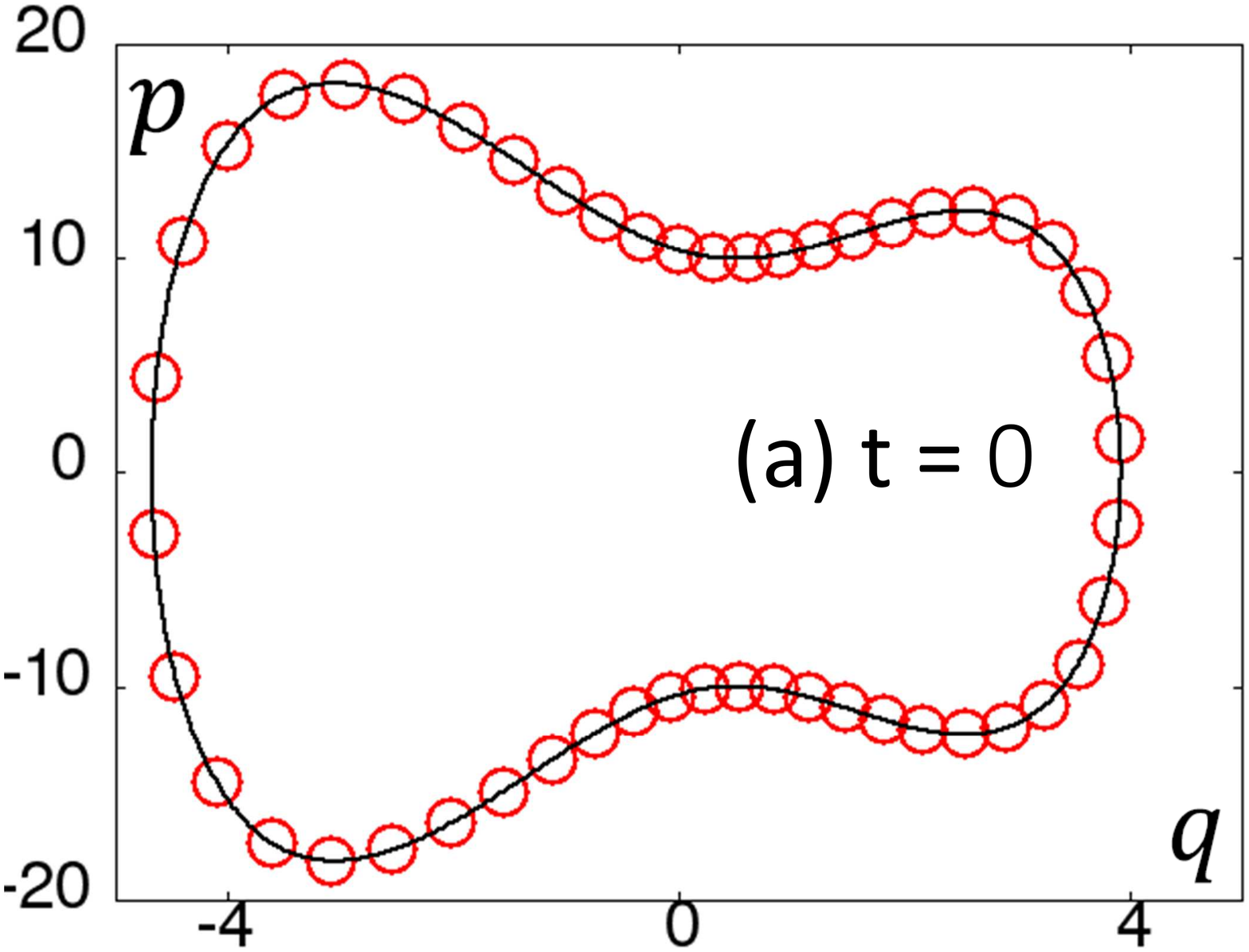}
      }
      \subfigure{
         \label{fig:class_interm}
         \includegraphics[trim = 0in 1in 2in 0in , scale=0.18,angle=0]{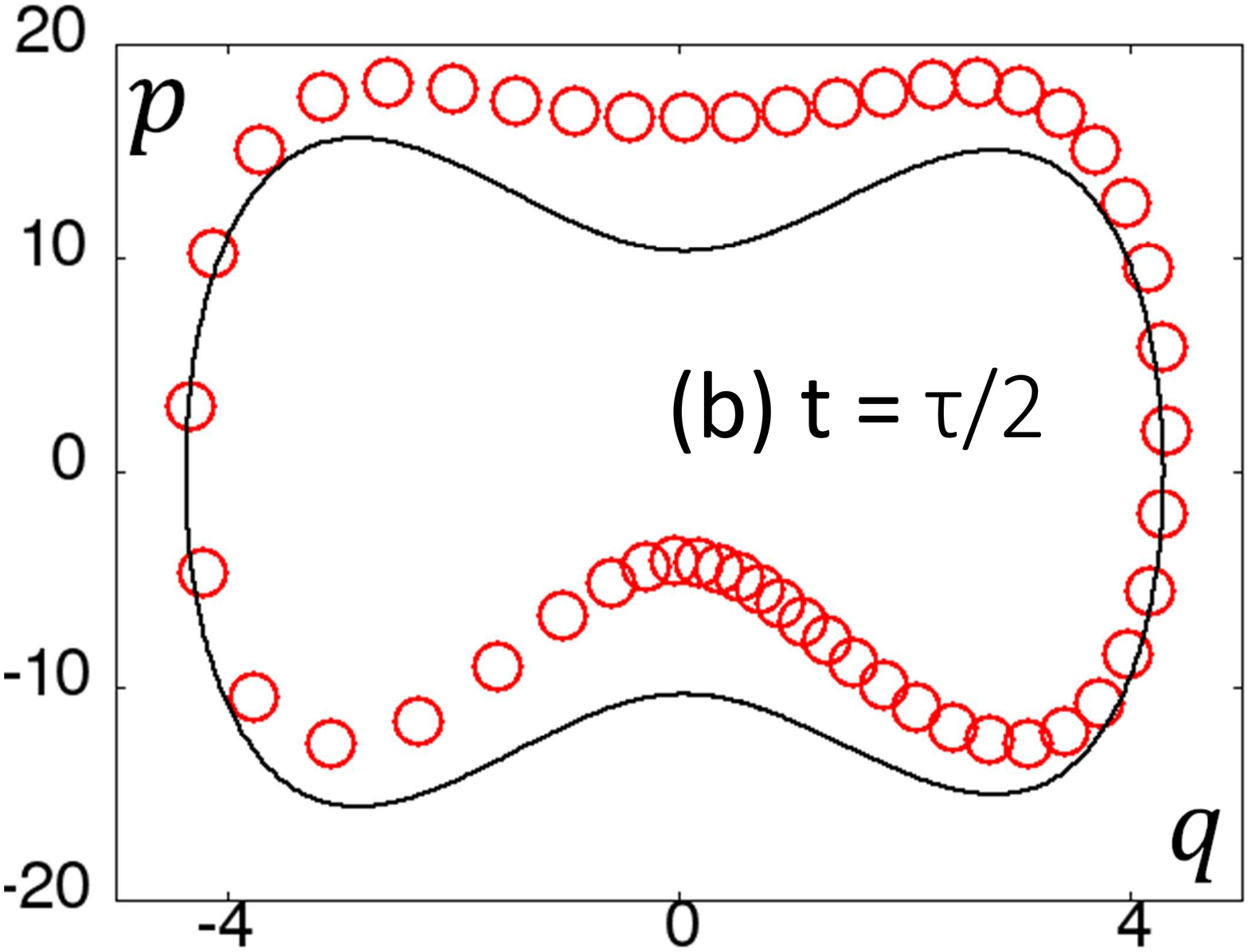}
      }
      \subfigure{
         \label{fig:class_final}
         \includegraphics[trim = 0in 1in 2in 0in , scale=0.18,angle=0]{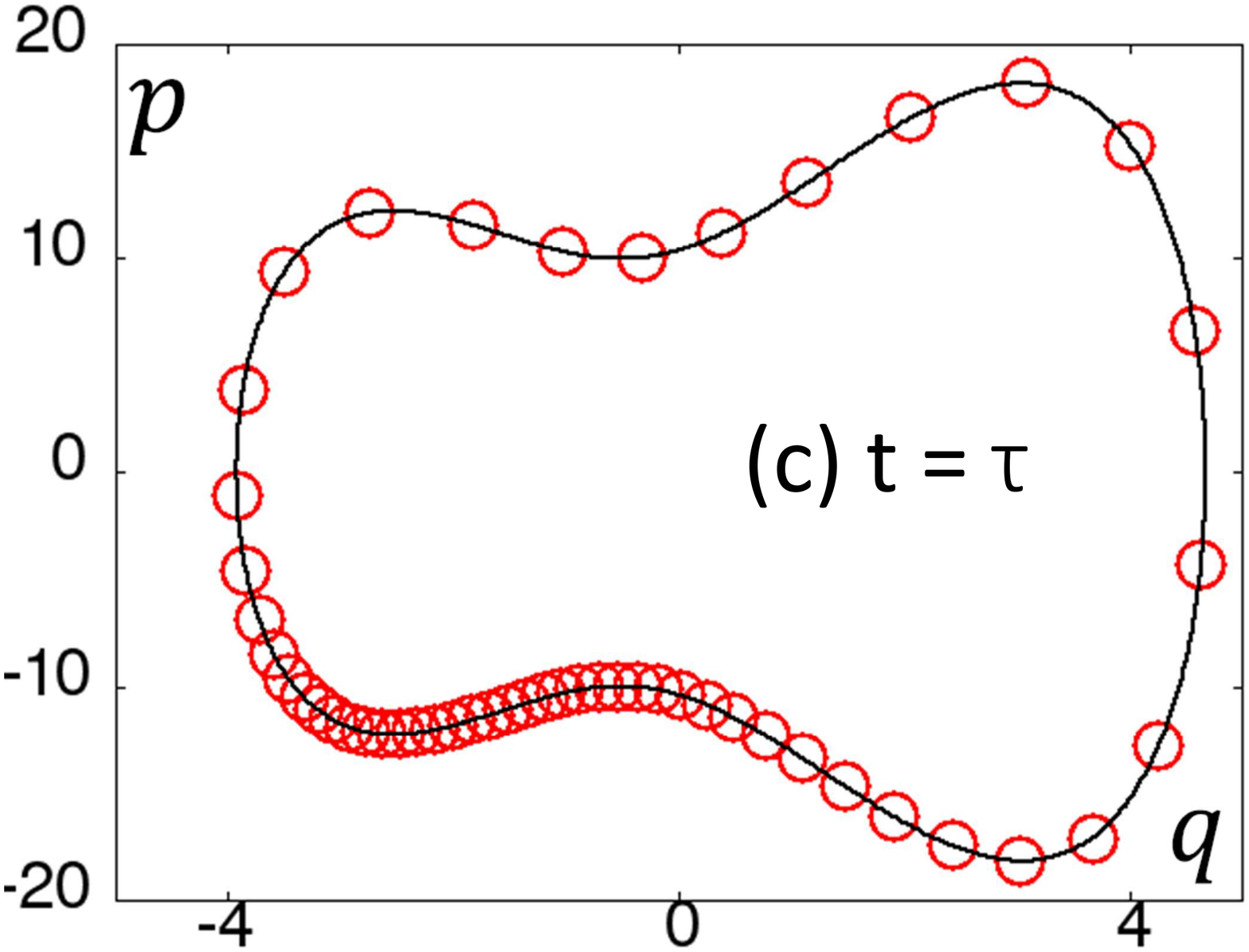}
      }
   \end{center}
   \caption{Snapshots of trajectories evolving in phase space under $H_0(q,p,t)+U_{FF}(q,t)$, the classical counterpart of the Hamiltonian used to generate Fig.~\ref{fig:prob_snapshot} (lower row).
   The closed black curves show the adiabatic energy shell ${\cal E}(t)$ of $H_0(q,p,t)$.
   }
      \label{fig:classical}
\end{figure}

Fig.~\ref{fig:classical} depicts snapshots of 50 trajectories evolving under the classical counterpart of the Hamiltonian $\hat H_0+\hat U_{FF}$ used to generate Figs.~\ref{fig:wcd_init} - \ref{fig:wcd_final} above, with $\tau = 1.0$.
Initial conditions were spaced uniformly, with respect to the angle $\theta$ of action-angle coordinates $(I,\theta)$~\cite{Goldstein80}, over an energy shell $H_0(q,p,0)=53.86=E_{17}(0)$.
The closed black curves depict an energy shell ${\cal E}(t)$ determined by a fixed value of the action, $I = (2\pi)^{-1} \oint \bar p \, dq$.
We refer to ${\cal E}(t)$ as the {\it adiabatic energy shell}.

The classical action $I$ is an adiabatic invariant:
for slow driving ($\tau\rightarrow\infty$), trajectories evolving under $H_0(q,p,t)$ 
cling to the adiabatic shell ${\cal E}(t)$ at all times,
whereas for rapid driving ($\tau=1.0$) the trajectories evolve away from the adiabatic shell~\cite{Jarzynski17}.
Fig.~\ref{fig:classical} illustrates the effect of adding the fast-forward potential $U_{FF}$ when $\tau=1.0$: the trajectories at first stray ``off shell'', Fig.~\ref{fig:class_interm}, but return faithfully to the adiabatic shell at $t=\tau$, Fig.~\ref{fig:class_final}.

Figs.~\ref{fig:class_init} - \ref{fig:class_final} are classical counterparts of Figs.~\ref{fig:wcd_init} - \ref{fig:wcd_final}.
Observe that both final states display ``imperfections'': the wavefunction in Fig.~\ref{fig:wcd_final} does not end exactly in the eigenstate $\phi_{17}(\tau)$, and the trajectories in Fig.~\ref{fig:class_final} are distributed non-uniformly over the energy shell ${\cal E}(\tau)$.
These features are related.
Using time-dependent WKB theory~\cite{Littlejohn1992}, we obtain the following prediction for the overlap between the final wavefunction $\psi(q,\tau)$ and the eigenstate $\phi_{n+l}(q,\tau)$ (see Appendix):
\be
\label{eq:fidelity}
\left\vert \bra \phi_{n+l} \vert \psi \ket \right\vert_{t=\tau}^2
= \left\vert \int_0^{2 \pi} d \theta \, e^{-i l \theta}  \sqrt{\frac{\eta(\theta,\tau)}{2 \pi}} \right\vert^2
\ee
where $\eta$ is the classical distribution of final conditions on the energy shell ${\cal E}(\tau)$, expressed in the angle variable $\theta$.
Eq.~\ref{eq:fidelity} relates the sidebands in Fig.~\ref{fig:wcd_final} to the non-uniformity of points in Fig.~\ref{fig:class_final} --
if the classical distribution were uniform, $\eta = 1/2\pi$, the sidebands would vanish: $\vert \bra \phi_{n+l} \vert \psi \ket \vert_{t=\tau}^2 = \delta_{l0}$.

\begin{figure}
\centering
\includegraphics[trim = 0in 1.8in 0in 1in , width=0.5\textwidth]{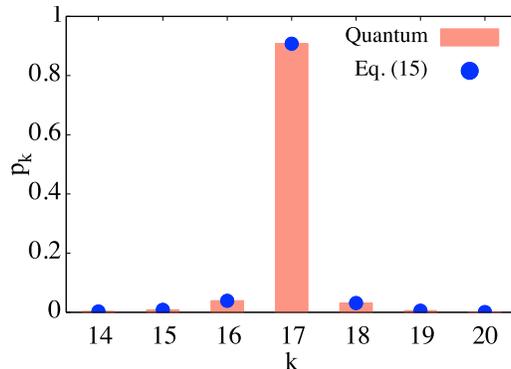}
\caption{Comparison between $p_k=|\bra \phi_k|\psi \ket|^2$ (red bars) obtained from quantum evolution under $\hat H_0+\hat U_{FF}$, and the prediction of Eq.~\ref{eq:fidelity} (blue dots) evaluated using the classical trajectories of Fig.~\ref{fig:classical}.}
\label{fig:Fourier}
\end{figure}

Fig.~\ref{fig:Fourier} compares the histogram appearing in Fig.~\ref{fig:wcd_final} with the prediction of Eq.~\ref{eq:fidelity}, evaluated using the trajectories in Fig~\ref{fig:class_final}.
The evident agreement supports our semiclassical analysis, which suggests that the non-zero sidebands will persist even in the limit $\hbar\rightarrow 0$.
Namely, imagine that for the same $\hat H_0(t)$ used to generate Fig.~\ref{fig:prob_snapshot}, we tuned the value of $\hbar$ (keeping other parameters unchanged) so that $E=53.86$ were the 1017'th, rather than 17'th, eigenenergy of $\hat H(0)$; and imagine we then re-ran both the quantum and classical simulations.
Since the latter are unaffected by the choice of $\hbar$, we would again obtain Fig.~\ref{fig:classical}, but the quantum simulations would produce a final histogram $p_k(\tau)$ sharply peaked at $k=1017$.
Eq.~\ref{eq:fidelity} implies that the distribution of probability among the eigenstates $k=1014-1020$ would be the same as that observed among $k=14-20$ in Figs.~\ref{fig:wcd_final} and \ref{fig:Fourier}.
This implies that in the limit $\hbar\rightarrow 0$, the variance of the energy distribution associated with the non-zero sidebands tends to zero (since the energy level spacing tends to zero), consistent with a classical final state in which all trajectories end on the adiabatic energy shell (Fig.~\ref{fig:class_final}).

\section{Conclusions}
\label{sec:conclusions}

We have expanded the toolkit for accelerating quantum dynamics, by developing a method for constructing fast-forward shortcuts that are free of divergences and thereby applicable to excited states of generic kinetic-plus-potential Hamiltonians in one degree of freedom.
Our approach is the semiclassical analogue of the classical method of Ref.~\cite{Jarzynski17}.
It will be interesting to extend our approach to systems with $d>1$ degrees of freedom.
For integrable systems the existence of WKB expressions for energy eigenstates are likely to prove useful to this end, but for systems with chaotic classical dynamics the task may prove challenging.
As noted elsewhere, in a number of experimentally relevant situations a separation of variables may effectively reduce a $d=3$ system to a $d=1$ system~\cite{Jarzynski17}.
Finally, while we have chosen to satisfy Eq.~\ref{cond_real} in our construction of $U_{FF}(q,t)$, and have used the Correspondence Principle to justify our choice, it will be interesting to test numerically whether satisfying Eq.~\ref{cond_im} instead would lead to a fast-forward potential capable of steering the wavefunction (approximately) to the desired final energy eigenstate.

From the start, we have assumed that the Hamiltonian $\hat H_0(t) = \hat p^2/2m + U_0(\hat q,t)$ is specified for all $t\in[0,\tau]$.
However, in other formulations of shortcuts to adiabaticity only initial and final Hamiltonians $\hat H_0(0)$ and $\hat H_0(\tau)$ are given, and the problem is then to construct a Hamiltonian $\hat H(t)$ under which an initial state $\psi(q,0) = \phi_n(q,0)$ evolves to a final state $\psi(q,\tau) = \phi_n(q,\tau)$.
In this situation, we can simply choose a potential $U_0(q,t)$ that interpolates smoothly from $U_0(q,0)$ to $U_0(q,\tau)$ and then apply our method to obtain $U_{FF}(q,t)$.
The Hamiltonian $\hat H(t) = \hat p^2/2m + U_0(\hat q,t) + U_{FF}(\hat q,t)$ then generates the desired evolution.

It is instructive to compare our results with those obtained for {\it scale-invariant} Hamiltonians~\cite{Deffner14}, which are characterized by time-dependence of the form
\be
\label{eq:scale-invart}
\hat H_0(t) = \frac{\hat p^2}{2m} + \frac{1}{\gamma^2} U_0\left(\frac{\hat q-f}{\gamma}\right)
\quad,\quad
f = f(t) ~,~ \gamma=\gamma(t)
\ee
Here, the parameters $f(t)$ and $\gamma(t)$ describe translations and dilations of the potential energy function.~\footnote{We note in passing that a harmonic oscillator described by the potential $k(t) [q-f(t)]^2/2$ falls into this class, with $U_0(x) = x^2/2$ and $\gamma = k^{-1/4}$.}
For Hamiltonians $\hat H_0(t)$ of this form, Eqs.~\ref{eq:continuity}, \ref{cond_real} and \ref{cond_im} are exactly satisfied by the non-divergent velocity field $v(q,t) = \dot f - (\dot\gamma/\gamma)(q-f)$, where dots indicate derivatives with respect to time.
Eq.~\ref{eq:aUFF} then leads to the fast-forward potential
\be
U_{FF}(q,t) = -m\ddot f q - \frac{m}{2} \frac{\ddot\gamma}{\gamma}(q-f)^2
\ee
The same potential was obtained in Ref.~\cite{Deffner14} using a trick involving canonical or unitary transformations of variables.
This comparison suggests that the success of that trick may be related to the fact that Eqs.~\ref{eq:continuity}, \ref{cond_real} and \ref{cond_im} are solved by a single, well-behaved field $v(q,t)$, which is a particular feature of scale-invariant Hamiltonians.
If this is the case then the approach of Ref.~\cite{Deffner14} might not readily generalize beyond scale-invariant Hamiltonians.
This issue deserves further investigation.

\begin{acknowledgments}
We gratefully acknowledge financial support from the U.S. Army Research Office under contract number W911NF-13-1-0390 (AP), and the U.S. National Science Foundation under grant DMR-1506969 (CJ). 
\end{acknowledgments}

\appendix

\setcounter{equation}{0}
\renewcommand{\theequation}{A\arabic{equation}}

\centerline{\bf Appendix}

Here we provide technical details of the derivations of Eqs.~\ref{eq:HJ}, \ref{eq:SED}, \ref{eq:continuity}, \ref{eq:perfect_cond} and \ref{eq:fidelity}.

\vskip .2in
\noindent\underline{\it Derivation of Eq.~\ref{eq:HJ}}
\vskip .1in

Suppose we have constructed a function $S^\circ(q,t)$ that satisfies $\partial_qS^\circ=mv$ (Eq.~\ref{eq:S}), without as yet adjusting the constant of integration $s(t)$.
From Eqs.~\ref{eq:aUFF} and \ref{eq:S} we obtain
\be
\partial_t \partial_q S^\circ = m\partial_t v = m (a - v\partial_q v) \\
= -\partial_q \left( U_{FF} + \frac{mv^2}{2} \right)
\ee
equivalently
\be
\partial_q \left[ \partial_t S^\circ + \frac{( \partial_q S^\circ )^2}{2m} + U_{FF} \right] = 0
\ee
hence
\be
\partial_t S^\circ + \frac{( \partial_q S^\circ )^2}{2m} + U_{FF} = \beta(t)
\ee
for some $\beta(t)$.
By setting
\be
S(q,t) = S^\circ(q,t) - \int_0^t \beta(t^\prime) dt^\prime
\ee
we arrive at $S(q,t)$ that satisfies both Eqs.~\ref{eq:S} and \ref{eq:HJ}.

\vskip .2in
\noindent\underline{\it Derivation of Eq.~\ref{eq:SED}}
\vskip .1in

Substituting $\psi(q,t)$ given by Eq.~\ref{eq:ansatz} into the time-dependent Schr\" odinger equation
\be
i\hbar \frac{\partial\psi}{\partial t} = \left( \hat H_0 + \hat U_{FF} \right) \psi
\ee
we separately evaluate each side using Eqs.~\ref{eq:H0} and \ref{eq:HJ} to obtain (dropping the subscript $n$)
\begin{subequations}
\label{eq:twosides}
\begin{align}
i\hbar \frac{\partial\psi}{\partial t}
= \left( i\hbar\dot\phi -\dot S \phi + E \phi \right) e^{iS/\hbar} e^{i\gamma}
= \left[ i\hbar\dot\phi + \left( \frac{S^{\prime\,2}}{2m} + U_{FF} + E \right) \phi \right] e^{iS/\hbar} e^{i\gamma} \\
\left( \hat H_0 + \hat U_{FF} \right) \psi
= \left[ -\frac{\hbar^2}{2m} \phi^{\prime\prime} - \frac{i\hbar}{2m}\left(S^{\prime\prime}\phi + 2S^\prime\phi^\prime\right)
+ \left( \frac{S^{\prime\,2}}{2m} + U_0 + U_{FF} \right) \phi \right] e^{iS/\hbar} e^{i\gamma}
\end{align}
\end{subequations}
where dots and primes denote $\partial_t$ and $\partial_q$, respectively, and $\gamma(t) = -(1/\hbar) \int_0^t E(t^\prime) dt^\prime$.
Setting the expressions in Eq.~\ref{eq:twosides} equal to one another and 
using $\hat H_0\phi=E\phi$,
we arrive at
\be
i\hbar\dot\phi = - \frac{i\hbar}{2m}\left(S^{\prime\prime}\phi + 2S^\prime\phi^\prime\right)
\ee
Using Eq.~\ref{eq:S} we then get
\be
i\hbar\dot\phi = - \frac{i\hbar}{2} \left( v^\prime \phi + 2 v \phi^\prime \right)
= - \frac{i\hbar}{2} \left[ \partial_q \left(v\phi\right) + v \partial_q\phi \right] =
\frac{1}{2} \left ( \hat p \hat v + \hat v \hat p \right) \phi
\ee
which is the desired result (Eq.~\ref{eq:SED}).

\vskip .2in
\noindent\underline{\it Derivation of Eq.~\ref{eq:continuity}}
\vskip .1in

Dividing both sides of Eq.~\ref{eq:SED} by $i\hbar$ and using $\hat p = -i\hbar\partial_q$, we get
\be
\frac{\partial\phi}{\partial t} = -\frac{1}{2\phi} \frac{\partial}{\partial q}\left(v\phi^2\right)
\ee
Multiplying both sides by $2\phi$ we obtain
\be
\frac{\partial \phi^2}{\partial t} = - \frac{\partial}{\partial q}\left(v\phi^2\right)
\ee
which is the continuity equation for the probability density $\phi^2(q,t)$.

\vskip .2in
\noindent\underline{\it Derivation of Eq.~\ref{eq:perfect_cond}}
\vskip .1in

Writing
\be
\phi_+(q,t) 
= \frac{\alpha}{\sqrt 2} \, A \, e^{i\Sigma/\hbar} 
\ee
where $A = \sqrt{\rho}$ and $\alpha=e^{-i\pi/4}$,
and substituting this expression into $i\hbar \partial_t \phi_+ = \hat D \phi_+$, the two sides evaluate as follows (dropping the extraneous factor $\alpha/\sqrt{2}$):
\be
\begin{split}
i\hbar \frac{\partial}{\partial t} \left( A e^{i\Sigma/\hbar} \right) &= \left( i\hbar \dot A - \dot\Sigma A \right) e^{i\Sigma/\hbar} \\
\hat D \left( A e^{i\Sigma/\hbar} \right) &= -\frac{i\hbar}{2} \left(
v^\prime A + 2 v A^\prime + 2vA \frac{i}{\hbar} \Sigma^\prime \right) e^{i\Sigma/\hbar}
\end{split}
\ee
Equating these two expressions, then dividing both sides of the equation by $e^{i\Sigma/\hbar}$, and then collecting real and imaginary terms, we obtain
\be
-\dot\Sigma A = v A \Sigma^\prime
\quad,\quad
\dot A = -\frac{1}{2} v^\prime A - vA^\prime
\ee
Dividing the first equation by $A$ and multiplying the second by $2A$, we arrive at Eq.~\ref{eq:perfect_cond}.
Evaluating $i\hbar \partial_t \phi_- = \hat D \phi_-$ produces identical results.

\vskip .2in
\noindent\underline{\it Derivation of Eq.~\ref{eq:fidelity}}
\vskip .1in

Time-dependent WKB theory~\cite{Littlejohn1992} provides a set of tools for constructing approximations to quantum wavefunctions evolving under the Schr\" odinger equation, from ensembles of classical trajectories obeying Hamiltonian dynamics.
Applying these tools to the final wavefunction $\psi(q,\tau)$ after evolution under $\hat H_0 + \hat U_{FF}$, we obtain (aside from an overall phase that we ignore)
\be
\label{eq:wkb-psiFinal}
\psi(q,\tau) = \sqrt{\eta_+} \, \alpha \, e^{i\Sigma_n/\hbar} +  \sqrt{\eta_-} \, \alpha^* \, e^{-i\Sigma_n/\hbar}
\ee
where $\alpha=e^{-i\pi/4}$, $\eta_\pm(q,\tau)$ denote the classical probability distributions on the upper and lower branches of the energy shell $E=E_n(\tau)$, depicted in Fig.~\ref{fig:class_final}, and $\Sigma_n(q,t)$ is given by Eq.~\ref{eq:Sigma}.
At a given location $q$, the two terms appearing on the right side of Eq.\ref{eq:wkb-psiFinal} describe a right-moving and a left-moving wave train, with local momenta $\pm\bar p = \pm \partial_q\Sigma_n$ (since the classical trajectories all end on the adiabatic energy shell) and amplitudes $\sqrt{\eta_\pm}$.

We wish to take the inner product between $\psi(q,\tau)$ and the $k$'th energy eigenstate $\phi_k(q,\tau)$, for $k\approx n$.
Time-{\it independent} WKB theory gives (see also Eq.~\ref{eq:wkb})
\be
\label{eq:wkb-revised}
\phi_k(q,\tau) = \sqrt{\rho_{k+}} \, \alpha \, e^{i\Sigma_k/\hbar} +  \sqrt{\rho_{k-}} \, \alpha^* \, e^{-i\Sigma_k/\hbar}
\ee
where $\rho_{k\pm} = \rho_k/2$ and the subscript $k$ indicates that we are evaluating $\Sigma(q,\tau)$ and $\rho(q,\tau)$ (Eq.~\ref{eq:Sigma}) at the energy $E=E_k(\tau)$.
The functions $\rho_{k\pm}$ represent equal contributions to the probability distribution $\rho_k$ from the two branches of the energy shell.

Taking the inner product between $\phi_k$ and $\psi$ gives the contribution
\be
\label{eq:innerProduct+}
\langle\phi_k\vert\psi\rangle_+ = \int_{q_L}^{q_R} dq \, \sqrt{\rho_{k+}\eta_+} \, e^{i(\Sigma_n-\Sigma_k)/\hbar}
= \int_{q_L}^{q_R} \rho_{k+} dq \, \sqrt{\frac{\eta_+}{\rho_{k+}}} \, e^{i(\Sigma_n-\Sigma_k)/\hbar}
\ee
from the upper branch, and a similar contribution $\langle\phi_k\vert\psi\rangle_-$ from the lower branch.
(We discard cross-term contributions between the two branches, as they approximately vanish upon integrating over rapidly oscillating phases.)
We now use action-angle coordinates $(\theta,I)$ to analyze these contributions.
In what follows, we mostly suppress the dependence of various quantities on the constant $\tau$.

The action $I$ corresponding to energy $E$ is given by
\be
I(E) = \frac{1}{2\pi} \oint_E \bar p \, dq \, .
\ee
Since the function $\Sigma$ is evaluated on a particular energy shell $E$, it can be viewed as a function of the action, $\Sigma = \Sigma(q,I)$, where $I=I(E)$.
Written in this way, it is the generating function for a canonical transformation from coordinates $(q,p)$ to $(\theta,I)$~\cite{Goldstein80}.
On the upper branch of the energy shell, the angle variable $\theta \in [0,\pi]$ is given by
\be
\theta(q,I) = \frac{\partial\Sigma}{\partial I}  \, .
\ee
Defining $I_j \equiv I(E_j) =  \hbar[j+(1/2)]$ we obtain
\be
\Sigma_n(q) - \Sigma_k(q) = \Sigma(q,I_n) - \Sigma(q,I_k) \approx \frac{\partial \Sigma}{\partial I} (I_n-I_k) = \theta \hbar(n-k) \, .
\ee

The distribution $\rho_{k+}(q)$ is uniform in the angle variable, $\rho_{k+}(q) dq = d\theta/2\pi$, which allows us to perform a change of variables, from $q$ to $\theta$, in the evaluation of Eq.~\ref{eq:innerProduct+}:
\be
\langle\phi_k\vert\psi\rangle_+ 
= \int_0^\pi \frac{d\theta}{2\pi} \, \sqrt{\frac{2\pi\eta_+}{\vert d\theta/dq\vert}} \, e^{i(n-k)\theta}
= \int_0^\pi d\theta \, \sqrt{\frac{\eta}{2\pi}} \, e^{i(n-k)\theta}
\ee
where
\be
\eta(\theta,\tau) = \eta_+(q,\tau) \left\vert \frac{\partial q}{\partial\theta} \right\vert \quad,\quad 0 < \theta < \pi
\ee
is the probability distribution on the upper branch of the energy shell, given in terms of the angle variable $\theta$.

For the lower branch of the energy shell we have $\theta = -\partial\Sigma/\partial I$ and $\rho_{k-} dq = -d\theta/2\pi$, and we obtain
\be
\langle\phi_k\vert\psi\rangle_- 
= \int_{q_L}^{q_R} \rho_{k-} dq \, \sqrt{\frac{\eta_-}{\rho_{k-}}} \, e^{-i(\Sigma_n-\Sigma_k)/\hbar}
= \int_{-\pi}^0 d\theta \, \sqrt{\frac{\eta}{2\pi}} \, e^{i(n-k)\theta}
\ee
with
\be
\eta(\theta,\tau) = \eta_-(q,\tau) \left\vert \frac{\partial q}{\partial\theta} \right\vert \quad,\quad -\pi < \theta < 0
\ee
Summing the two contributions, we get
\be
\langle\phi_k\vert\psi\rangle = \int_{-\pi}^\pi d\theta \, \sqrt{\frac{\eta(\theta,\tau)}{2\pi}} \, e^{i(n-k)\theta}
\ee
Replacing $\int_{-\pi}^\pi d\theta$ with $\int_0^{2\pi} d\theta$ and setting $l = k-n$, we arrive at Eq.~\ref{eq:fidelity}.

\begin{center}
\line(1,0){400}
\end{center}

\vskip .1in
\noindent
\underline{Animated .gif files:}
\vskip .1in

\noindent
$\bullet$ {\it movie\_withoutFastForwardPotential.gif} : shows $p_k(t) = \vert \left\langle \phi_k(t)\vert\psi(t)\right\rangle\vert^2$ plotted as a function of $k$ (red bars), for a quantum system evolving in the potential given by Eq.~\ref{eq:modelHamiltonian} of the main text.  Parameters are the same as for Fig.~\ref{fig:prob_snapshot}, except $\hbar = 1$ and $\psi_0 = \vert 35 \rangle$.
The inset shows $U_0(q,t)$.

\vskip .1in
\noindent
$\bullet$ {\it movie\_withFastForwardPotential.gif} : same as above, but with the addition of $U_{FF}(q,t)$.
The inset shows $U_0(q,t)$ and $U_{FF}(q,t)$.

\bibliography{References}

\end{document}